\documentclass{llncs}

\usepackage{amsmath}
\usepackage{hyperref}

\usepackage[auth-lg]{authblk}
\usepackage{bookmark}
\usepackage{booktabs}
\usepackage[final]{listings}
\usepackage{lscape}
\usepackage{mathtools}
\usepackage{paralist}
\usepackage{flushend}
\usepackage{changebar}

\usepackage{natbib-cite}
\usepackage{natbib-acm}

\usepackage{hypernat}
\usepackage{cleveref}

\crefname{section}{Section}{Sections}
\crefname{table}{Table}{Tables}
\crefname{figure}{Figure}{Figures}
\crefname{subfigure}{Figure}{Figures}

\begin{document}

\title{Information-Gain Computation}

\author{Anthony Di Franco}
\institute{Department of Computer Science\\
University of California, Davis\\
Davis, CA 95616\\
\email{acdifranco@ucdavis.edu}}

\maketitle

\begin{abstract}

Despite large incentives, correctness in software remains an elusive goal.
Declarative programming techniques, where algorithms are derived from a
specification of the desired behavior, offer hope to address this problem, since
there is a combinatorial reduction in complexity in programming in terms of
specifications instead of algorithms, and arbitrary desired properties can be
expressed and enforced in specifications directly.

However, limitations on performance have prevented programming with declarative
specifications from becoming a mainstream technique for general-purpose
programming. To address the performance bottleneck in deriving an algorithm from
a specification, I propose information-gain computation, a framework where an
adaptive evaluation strategy is used to efficiently perform a search which
derives algorithms that provide information about a query via the most efficient
routes. Within this framework, opportunities to compress the search space
present themselves, which suggest that information-theoretic bounds on the
performance of such a system might be articulated and a system designed to
achieve them.

In a preliminary empirical study of adaptive evaluation for a simple test
program, the evaluation strategy adapts successfully to evaluate a query
efficiently.

\end{abstract}

\section{Introduction and Motivation}

Despite large incentives, correctness in software remains an elusive goal.
Declarative programming techniques, where algorithms are derived from a
specification of the desired behavior, offer hope to address this problem, since
there is a combinatorial reduction in complexity in programming in terms of
specifications instead of algorithms, and arbitrary desired properties can be
expressed and enforced in specifications directly. Additionally, giving an
explicit specification preserves information about program semantics and
programmer intent that is lost by forcing the programmer to manually translate
an explicit or implicit specification into an algorithm that implicitly and
usually only partially satisfies that specification, information that may be
used by automated systems to implement correct behavior in a performant way.

However, limitations on performance have prevented programming with declarative
specifications from becoming a mainstream technique for general-purpose
programming. Without domain-specific knowledge, default evaluation strategies
must strike a sophisticated balance among efficiency, the semantic properties of
soundness and completeness, and simplicity, which is relevant both to
implementation effort and to comprehensibility by the programmer, if the
programmer must be relied upon to implicitly influence the behavior of the
search to achieve efficiency, which is in practice how efficiency is achieved in
general-purpose declarative languages such as Prolog. Furthermore, because of
the combinatorial nature of the searches involved in evaluation, exponential or
worse reductions in efficiency can result from deviations from the best
evaluation strategy, meaning these efficiency concerns are often decisive in
whether a program is practical to use at all; they are concerns of the highest
order. This, of course, undermines the status of such languages as declarative,
since the task of programming still involves understanding and influencing the
evaluation of the program at an algorithmic level, and it is insufficient to
program only in terms of the declarative semantics of the problem.

Kowalski's framing of algorithm = logic + control in the paper of the same
name\citep{Kowalski:1979:ALC:359131.359136} provides guidance here. Accordingly,
to obtain an algorithm that implements the logic of a specification, we must add
a control component (that is, a choice of order of evaluation of the declarative
logic) that uses that logic to produce the desired result efficiently.

To address the performance bottleneck in deriving an algorithm from a
specification in this way, I propose \textit{information-gain computation}, a
framework where an adaptive evaluation strategy is used to efficiently perform a
search which derives algorithms that provide information about a query most
directly. The key aspect is to measure information gain about a goal when a
certain control choice is made in a certain context, and adapt these choices to
increase the rate of information gain. Information gain provides a meaningful
measure of progress for a computation, which in turn provides an objective for
optimization for the adaptive strategy. Measuring the progress of a computation
in general has proven difficult because unbounded effort can be expended without
any indication as to whether the computation will halt if allowed to continue.
Here, halting is replaced with yielding information, and execution proceeds
nondeterministically to find paths that yield information at the highest rate,
sidestepping these issues. Also, the factoring of the program logic into
recursive predicates makes it possible to share information about effective
control choices to achieve statistical efficiency.

Within this framework, opportunities to compress the search space present
themselves, first, by identifying the traces that contribute the most
information to answering a given query, and then, by compressing these traces.
This framing suggests that information-theoretic bounds on the performance of
such a system might be articulated and a system designed to achieve them,
holding promise for a definitive solution to the problem of deriving an
algorithm from a specification, and thus to that of declarative software.

To test the core idea, adaptation of the evaluation strategy with respect to
information gain, adaptive evaluation of a fixed Prolog program was implemented
and the efficiency measured and the induced control structure described. The
program was adapted from an elementary Prolog programming example for beginners
where ancestorship is computed recursively in a directed graph. The program was
modified from its usual form to introduce an extremely large amount of sparsity
in the search space, but in a form where sharing of statistical strength due to
the explicitly recursive structure of the description of the search space should
permit quickly learning a bias away from the costly diversions. The program
confounds Prolog's default evaluation strategy even on problems of trivial size
but the adaptive evaluation strategy succeeds in evaluating the program
efficiently.

The rest of this paper is organized as follows. Section 2 presents a focused
review of related work. In Section 3, I propose an architecture for a system
implementing information-gain-driven evaluation. Section 4 describes the design
of an experiment which uses an adaptive evaluation strategy on a test program.
Section 5 discusses the results of the experiments. Section 6 presents
conclusions and discusses future work.

\section{Related Work}

Prolog is the most well-established and widely-used logic programming language.
Its recursive, top-down decomposition of queries (goals) into subgoals will be
retained and used to advantage here.\citep{ISO:1995:IIIe}

Blog\citep{Li:EECS-2013-51} and Problog\citep{Raedt07problog:a} are examples of
languages that combine logical and probabilistic semantics, however they do not
attempt to take advantage of this to improve their evaluation strategies or
realize true declarative semantics, instead relying on standard logic program
evaluation and extending it with a variety of ad-hoc techniques for
probabilistic portions of the program. To the best of my knowledge, using
probabilistic semantics to achieve efficient evaluation in a general-purpose
setting is a novel approach, and it yields benefits for purely logical programs
as well as advantageously framing the problem of integrating probabilistic
information into a logical framework.

Schmidhuber's adaptive history compression\citep{Schmidhuber91adaptivehistory}
will form the basis for the proposed trace compression, along with
Info-clustering\citep{DBLP:journals/corr/ChanAZKL16}, both of which will be
discussed in that context.

Bandit algorithms will play the central role in adapting the evaluation
strategy. Bandit algorithms address the problem of balancing exploration and
exploitation in choosing actions with unknown rewards; this is the problem we
face in trying to choose a branch to explore in evaluation of a program in hopes
of gaining information. The literature on bandit algorithms is vast; a highlight
relevant to the present and future work here is Thompson sampling, a Bayesian
framework\citep{Ortega_aminimum}. The present work uses \texttt{UCB1}, a simple
rule that orchestrates exploration and exploitation according to a bound on an
expected reward estimate modified with an additive uncertainty term which
decreases as a choice is sampled according to a simple error model. In Powley
and Cowling's work\citep{Powley_banditsall} there is some precedent for using
\texttt{UCB1} in a similar setting, to explore an unbounded tree that yields
reward only at its leaves.

\section{Design for Information-Driven Evaluation}

This section describes an architecture for a system to perform adaptive
evaluation and suggests a strategy for a concrete implementation.

The proposed design has three main aspects. First, an adaptive technique is used
to identify control choices that tend to yield the most information about the
query given the data. Second, the results of executing these control choices are
examined to determine the most frequently executed traces (equivalently,
information propagation paths from data to the query). Third, these traces are
recursively compressed in a way that creates new, shorter information
propagation paths from the data to the query.

Before considering the design we begin with a review of the aspects of the usual
evaluation strategy in a relational language with emphasis on the features that
lead to difficulties the current design addresses.

\subsection{Review of Evaluation in Relational Languages}

In a relational language such as Prolog, a program consists of rules that refer
to other rules recursively, and which ultimately may match facts (data).
Evaluation proceeds from a \textit{query}, which is a top-level goal that is
recursively expanded to subgoals in a left-to-right, depth-first search. When a
fact is matched, it is added to an answer set for the goal it matches, and these
sets are combined according to the logic of the predicates of the goals they
appear in until they produce a final answer set for the query.

Depth-first search requires the least state for a search; only a stack of
previous goals and position in them must be maintained. However, one drawback of
this approach is that it is incomplete; for example, a left-recursive rule would
prevent the search from terminating since the rule would expand into itself ad
infinitum before any other branch is taken.

Each predicate thus implicitly represents a discrete joint space of variation,
and, given data, we can construct the joint spaces they represent explicitly by
evaluating the program.

The Prolog strategy described above is a \textit{backward-chaining} strategy; it
is also possible to use \textit{forward-chaining}, which applies rules in a
bottom-up fashion, starting from facts that match rules and then recursively
invoking rules those rules appear in as subgoals until the query is reached. The
identification and use of traces described below uses adaptive backward-chaining
to identify when forward-chaining would be advantageous and apply it in those
cases.

\subsection{Adaptive Evaluation With Bandit Algorithms}

As mentioned in the introduction, we take the information about the query as the
objective of evaluation, or rather more specifically the information gained per
unit of computational effort expended, and treat the choice of which subgoal to
evaluate in each goal as a \textit{multi-armed bandit} problem.\citep{CIS-103707}
Briefly, in each goal we face a choice of which subgoal's tree to explore, in
hopes it will match and yield information about facts at some point. Whenever we
encounter subgoals that correspond to the same predicate, we can share
information about subgoal choice, since each predicate is identically an
implicit representation of its own joint membership space.

To measure progress in gaining information about a goal, we measure the decrease
in the \textit{total correlation}\citep{Watanabe:1960:ITA:1661258.1661265}, which
is the sum of the entropies of the individual variables minus the entropy of
their joint distribution. For random variables $X_1,X_2,\ldots,X_n$, the
expression is this: $$\sum_{i=1}^n H(X_i) - H(X_1, X_2, \ldots, X_n)$$ This is
equivalent to the Kullback-Leibler divergence from the joint distribution to the
independent distribution of the variables. In the set of variables in this
calculation, we include the variables appearing in the goal, as well as a
variable representing a prior over the joint space from which the variables in
the goal may be drawn. Thus under this measure, information is gained whenever
variables are distinguished from one another or when the joint distribution is
distinguished from the prior. For logic programming, we may use the uniform
prior over the discrete set of facts.

Algorithms for multi-armed bandit problems admit several desiderata. Those
relevant here are succeeding with high probability vs. only in expectation, and
contextuality, or the ability to condition decisions on side information.
The ability to succeed under an adversarial choice of rewards rather than only
with i.i.d. rewards may be important in problems that closely model an
adversarial setting.

The ability to take context into account can be used to take the current state
of information about the goal into account. The choice of branch to explore next
can be conditioned on the amount of effort spent exploring each branch so far,
or a summary statistic thereof, together with any contextual restrictions on the
search space resulting from top-down unification, because together they are a
sufficient statistic for the current state of knowledge. It may also be possible
to condition on the current state of knowledge directly, for example by
representing the joint distribution representing the current beliefs in an
efficient basis and conditioning on its parameters.

Bearing in mind these desiderata, we can proceed with
\texttt{Exp4.P},\citep{Beygelzimer2011ContextualBA} an algorithm which works in
both stochastic and adversarial conditions, accepts contextual information,
succeeds with high probability, and achieves a regret bound with a square root
of a log factor of the optimal bound.

We associate the state of a bandit choice algorithm with each predicate in the
program, and measure the information gained per unit effort in exploring a
branch, using a budget for further expansions that may occur during that
exploration to ensure that the exploration terminates and the effort can be
accounted for properly. This budget may be incrementally increased as in an
iterative-deepening search strategy.

\subsection{Hot Traces and Optimistic Forward-Chaining}

The most frequently taken control paths in evaluating the program are now
determined by a near-optimal procedure for selecting those that yield the most
information with the least computational effort, which we may view as a first,
most basic optimization in evaluating the program. We can identify these control
paths either by sampling control stacks to gather statistics during execution or
by examining the weights of the bandit choice algorithms for a query and its
constituent subgoals.

Identifying these traces brings several benefits. First, they can be
optimistically executed in forward-chaining mode whenever a fact that matches
one is known, since they have already been determined to belong to a
near-optimal algorithm for taking that fact into account. Second, they may be
heavily optimized by techniques from tracing JITs, being inherentlty in the
appropriate form. Third, since they are lacking in control flow and specify the
type of data they consume explicitly, they are especially amenable to execution
on efficient dataflow-oriented hardware such as GPUs, FPGAs, and vector
processors.

Another potential benefit, perhaps the most significant one because it holds the
promise of a tight bound on the effort required to answer the query, is
compression of the search/inference space, described below.

\subsection{Compressing Traces}

We can attempt to compress traces in the following way, a generalization of
Schmidhuber's history compression\citep{Schmidhuber91adaptivehistory} for
sequences which does not require a total order, i.e. to space-like rather than
time-like relations.

In history compression, a hierarchy of representations of a sequence is learned
by learning a predictor of the sequence, and constructing a new sequence that
consists of the indices of mispredicted symbols along with the correct symbol,
and so on recursively with that sequence as desired or until no more compression
is obtained.

To generalize beyond sequences, I propose using
\textit{info-clustering}\citep{DBLP:journals/corr/ChanAZKL16} to learn dependency
structures within each predicate's empirical joint distribution, and to learn a
probabilistic model of each of those joint distributions. The parameters of
these models, along with their residual errors, are analogous to the models and
their mispredictions used by history compression. Now, to build up a recursive
hierarchy of representations as in history compression, we take the model
parameters and residual errors and consider the joint spaces of those which
appear adjacent to one another on a trace, and recursively apply info-clustering
and joint-space modeling on those until no more compression can be obtained. The
result is a tree of reduced descriptions of the joint spaces, which may provide
exponentially shorter information propagation paths from facts to query. Facts
entering can be transformed through a number of models logarithmic in the length
of the trace to yield information about the query efficiently.

\section{Experiments}

\subsection{Simulating Adaptive Evaluation}

Due to time restrictions, modifying existing Prolog implementations and
implementing a simple Prolog-like language from scratch proved prohibitively
complex and involved many aspects outside the scope of the present study, so to
investigate the fundamental feasibility of the strategy arising from the framing
here we simulate the relevant aspects of the search for solutions that would be
used by a relational language with the proposed adaptive evaluation strategy. To
simulate evaluation, we fix a program and data and implement by hand the search
that corresponds to applying the evaluation strategy to that program and data.

\subsection{Example Preliminaries}

As an elementary example, consider the following code to compute the transitive
closure of a graph (phrased in terms of the elementary Prolog programming
example about recursively finding ancestors of a person given a set of
parent-child relationships):

\begin{lstlisting}
 ancestor(A, B) :- parent(A, B).
 ancestor(A, B) :- parent(A, X),
                   ancestor(X, B).
\end{lstlisting}

That is, A is an ancestor of B either if A is a parent of B, or if A is a parent
of another X and X is an ancestor of B.

We can use this as the basis of a simple example to test the effects of an
adaptive evaluation strategy. To represent the effects of sparsity of the
search space, which is the main obstacle that adaptive evaluation is meant to
address, we add additional rules that confound the search, like so:

\begin{lstlisting}
 ancestor(A, B)   :- parent(A, B).
 ancestor(A, B)   :- deadend(A, B).
 deadend(A, B)    :- deadend(A, B,
                       100000000).
 deadend(A, B, N) :- N1 is N - 1, N > 0
                     -> deadend(A, B, N1);
                        fail.
 ancestor(A, B)   :- parent(A, X),
                     ancestor(X, B).
\end{lstlisting}

Further we assume these data:

\begin{lstlisting}
 parent(tom, fred).
 parent(fred, jill).
\end{lstlisting}

Under the default Prolog evaluation strategy, the deadend rule is preferred to
the informative rules for pursuing the ancestor search, because it appears first
in the program text, resulting in the program counting down from 100 million
before resuming a productive branch of the search, and causing a delay of about
10 seconds to answer the query \texttt{?- ancestor(tom, jill).} on SWI-Prolog
7.4.2 on a 2.8 GHz Intel Xeon CPU E3-1505M v5. An adaptive search should be able
to detect that a great deal of work is being performed without yielding
information on this branch and should thereafter strongly prefer another.

\subsection{Implementation}

This Prolog program under an evaluation strategy that uses
\texttt{UCB1}\citep{Powley_banditsall} (for the sake of simplicity of
implementation) to choose each subgoal was implemented in Python as a program
that selects a branch to explore with \texttt{UCB1} in a loop and updates the
answer set and \texttt{UCB1} parameters as facts are encountered. Two sets of
benchmark data are used: one corresponding to the Tom, Fred, Jill example above,
and one that creates a 1000x1000 upper-triangular matrix where elements above
the diagonal are 1 with $p=1/8$ for a 1 entry, $p=7/8$ for a 0 to represent
parenthood. The query in this case asks for the indices of nodes with node 0 as
an ancestor. The information accounting ignores structure in the joint space and
assumes the uniform over pairs in the joint space as the prior, gaining a bit of
information whenever a descendant is identified, and neglecting to check when it
becomes known that an element cannot possibly be a descendant. The time,
\texttt{UCB1} expected information gain estimates, and number of times each
predicate was branched to were reported.

\section{Evaluation}

For the Tom, Fred, and Jill example, a negligible amount of time is taken, and
the deadend branch is taken once (which, however, requires it to be played
100,000,000 times because it recurs only into itself that many times,) while the
informative branch is taken 3 times, to yield the correct answer. The
\texttt{UCB1} weights learned were about 0.0006 for the deadend branch and 3.037
for the informative branch, showing that approximately all 2 bits of information
were estimated to have come from the informative branch, which was the case.
Since reward was about one bit per branch down the informative branch, the much
higher weight showed that in only 3 searches of that branch, the uncertainty
bound could not be improved to even within the same order of magnitude as the
true expected utility value, and the presence of the number 3 in both the
leading digit of the branch weight and the number of times it was taken is a
coincidence.

For the 1000x1000 random matrix example, execution took about 40 seconds, and
again took the deadend branch once by choice and 100,000,000 times total,
thereafter taking the informative branch 145,205 times and again learning a
weight of about 0.0006 for the deadend branch but a weight of about 1.0092 for
the informative branch, which is very close to the true expected value of one
bit per branch taken, the uncertainty bound penalty in the weight term having
been reduced to a negligible level by the large number of tries of the branch.

\section{Conclusions, Comments, and Future Work}

The empirical test of the bandit-algorithm-driven evaluation branch choice
proved successful, even with an example program that can be made to take
arbitrarily long in Prolog with a choice of the parameter setting the depth of
the pathological branch.

This suggests that the additional work to build information-gain-driven adaptive
evaluation into a relational language is justified. For a pure-logic language,
the technique of defaulting to uniform prior over a discrete domain of facts
used in the simulated evaluation here may suffice, but to use non-uniform
priors, to represent correlations and information-sharing in the joint spaces in
predicates properly, and to move forward with ideas for search space
compression, a full probabilistic-relational language is called for.

So far, only declarative computation - lacking in side effects - has been
considered. To embrace interaction with an environment with mutable state into
the same framework in a way that preserves its strengths, we can associate
program fragments with predicates, and thread program fragments together along
traces to produce plans as in Pyke, a Prolog-like relational language
implemented in Python.\citep{Pyke} This will introduce additional concerns in
working with compressed traces.

The potential for information-theoretic optimality of the evaluation strategy
induced by the proposed methods was mentioned in passing but not discussed in
toto. The claim is that the bandit algorithm generates efficient evaluation
plans, and that relatively few traces out of all possible traces will contribute
the majority of the obtainable information. Then, compression of these few
traces in such a way that new information propagation paths of length
proportional to the information content of the joint relation space along the
entire trace will reduce evaluation effort along these traces to costs of the
order of an information-theoretic bound. Practical difficulties may arise in the
expense of transforming through the hierarchy of models. Likewise, practical
difficulties in optimizing and parallelizing traces may arise in residual
control components inherent in the semantics of the individual base predicates
in the system, and in accounting for interactions with side-effecting plans as
mentioned in the previous paragraph.

In a probabilistic-relational language that had implemented the recursive
joint-space modeling in trace compression, data could be supplied with an
uninformative predicate structure, similar to that used in a convolutional
neural network, where the same relationship is assumed in each local
neighborhood in a larger joint space. The trace compression would then
automatically build a hierarchy of autoassociators to model the data as they
relate to the query. This suggests computational interpretations of
neuroanatomical structures that may be explored in future work - cortex as a
hierarchy of autoassociating compressors attempting to learn short information
propagation paths from senses to query-like structures directing evaluation in
the basal ganglia, emitting motor plans along these traces to effect attention
and interaction with the environment, and optimizing motor plans and maximizing
their dependence directly on data in the cerebellum.

For choices with large numbers of arms (such as for selection of a discretized
continuous parameter,) the \texttt{CEMAB} method\citep{Wang2017} may be
applicable. It is derived from the cross-entropy method, an importance-sampling
technique for rare-event simulation later adapted to hard optimization problems.

\bibliography{dblp}

\end{document}